\documentstyle[sprocl]{article}

\input{epsf}

\newcommand{\lae}{\mathrel{<\kern-1.0em\lower0.9ex\hbox{$\sim$}}}
\newcommand{\gae}{\mathrel{>\kern-1.0em\lower0.9ex\hbox{$\sim$}}}
\def\Journal#1#2#3#4{{#1} {\bf #2}, #3 (#4)}
 
\begin{document}

\title{PHOTOMETRY AND SPECTROSCOPY OF COMA DWARF ELLIPTICALS}

\author{JEFF SECKER}
\address{Program in Astronomy, Washington State University, Pullman, WA 99164-3113 USA}

\author{WILLIAM E. HARRIS}
\address{Department of Physics and Astronomy, McMaster University, Hamilton, Ontario L8S 4M1 Canada}

\author{PAT C\^OT\'E and J.B. OKE}
\address{California Institute of Technology, Pasadena, CA 91125 USA, \\
and \\
Dominion Astrophysical Observatory, National Research Council, 5071 W. Saanich Rd.,  Victoria, BC V8X 4M6 Canada}


\maketitle

\section{Destruction of Faint Dwarf Ellipticals in the Core}

We have used deep $B-$ and $R-$band CCD images (with $\simeq 1.2^{''}$
seeing) of the central $\simeq 700$ arcmin$^2$ of the Coma cluster
core (and an associated $\simeq 270$ arcmin$^2$ control field) to
study the early-type dwarf elliptical (dE) galaxy population (Secker
\& Harris 1997; Secker, Harris \& Plummer 1997).
In the left panel of Fig. 1, we plot a subset of the total cluster
field CMD: the galaxy sequence is clearly visible, with the dEs
beginning at $R \simeq 15.5$ mag and a mean color of
$(B-R) \simeq 1.54$ mag.  The larger solid circles represent the
trimmed median $(B-R)$ color in one-magnitude bins over the entire
luminosity range, and the solid line illustrates the least-squares
regression line fit to median color values in the range
$14 < R < 18.5$ mag.  This provides an excellent fit, and the bright
part of the early-type galaxy sequence shows a strong trend for
fainter dE galaxies to have (on average) a bluer color: $\Delta (B-R)
/ \Delta R = -0.056\pm0.002$.  Fainter than $R \simeq 18.5$ mag, the dE
sequence spreads and merges with the multitude of noncluster galaxies,
which are the dominant population at 
these magnitudes.  The effect is that the program-field mean color
values are skewed redward, away from the dashed line (an extension of
the upper solid line), towards the mean colors of the control-field
objects (open circles).

\begin{figure}[ht]
\epsfysize=3.0in
\epsfbox[60 250 380 560 ]{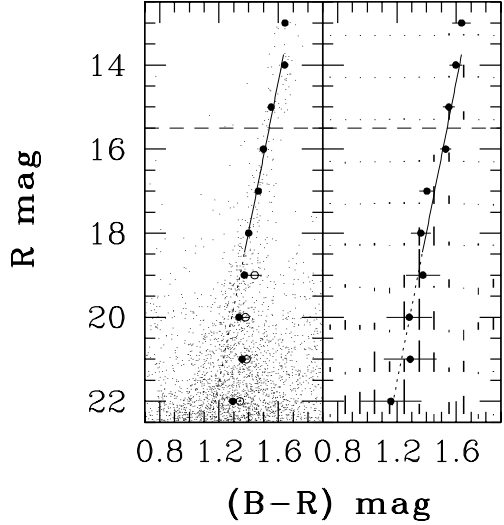}
\caption{Analysis of the control-field corrected program-field CMD.}
\end{figure}

In the right panel of Fig. 1, we plot the binned, scaled and
subtracted CMD: one vertical line segment for each color-magnitude
bin, with a height proportional to the number of objects in that bin.
Here, the solid circles represent the mean corrected color, weighted
by the number of objects in each bin.  The regression line from the
left panel is reproduced here, together with the extrapolated dotted
line: {\em it provides an excellent fit to the corrected mean color
values of the dE galaxy sequence, over the full range $15.5 \le R \le
22.5$ mag}.  Thus the apparent reddening of the dE sequence in the
left panel is attributed entirely to contamination of the cluster
sample.

Using the color-metallicity relationship derived from metallicities
and integrated colors for Galactic globular clusters, valid over the
range $-2.4 \lae$ [Fe/H] $\lae -0.2$ (Harris 1996), we obtain
$\Delta$[Fe/H]$/ \Delta R = -0.19\pm0.01$ dex for our sample of Coma
dEs. Cellone, Forte \& Geisler (1994) plot a $T_1$, $(C-T_1)$ CMD
(Washington filters; $T_1 \simeq R$) for a sample of 14 Fornax dEs.
From a linear fit to their tabulated data points, and the
color-metallicity relationship of Geisler \& Forte (1990), we
calculate $\Delta$[Fe/H]$/ \Delta T_1 = -0.18\pm0.07$ dex for these
Fornax dEs.  These slopes are formally consistent, suggesting that the
process of dE galaxy formation is very similar in these very
different cluster environments.

We have also analyzed the completeness-corrected background-subtracted
radial number density profiles of bright and faint galaxies in the
cluster core.  Our bright sample includes 280 cluster galaxies, a
subsample selected to have $0.7 \leq (B-R) \leq 1.9$ mag and $R \leq
19.0$ mag (giants plus bright dEs).  Our faint sample includes 2246
objects selected to have $0.7 \leq (B-R) \leq 1.9$ mag and $19.0 < R
\le 22.5$ mag.  For the bright galaxy sample, the best-fit King model
has a core radius $R_{\rm c} = 13.7$ arcmin, significantly smaller
than $R_{\rm c} = 22.2$ arcmin found for the faint dEs.  Lobo et
al. (1997) analyzed the faint-end slope of the galaxy luminosity
function in Coma, and determined that it is shallower in the core than
for the cluster as a whole.  This is consistent with the dynamical
destruction of dE galaxies in the dense environment of the cluster
core.  Our results presented above are also consistent with this
hypothesis:

(1) Thompson \& Gregory (1993) proposed that the core of the Coma
cluster is deficient in the number of faint dE galaxies, and they
argued that tidal disruption could be responsible for partially
destroying this low-mass population.  Our study indicates that the
larger core radius for faint dEs is genuine, and in this sense
the cluster core is deficient in the number of faint dEs.

(2) Mattila (1977) measured the color of the diffuse intergalactic
component of the Coma cluster light, and obtained $(B-V) =0.54\pm0.18$
mag.  From the right panel of Fig. 1, we estimate $(B-R) \simeq 1.15$
mag for our faintest dEs ($R = 22.5$ mag), corresponding to $(B-V)
\simeq 0.70$ mag, which is within the color range found by Mattila
(1977).  This is consistent with a major component of this diffuse
intergalactic light originating as stars tidally stripped from
numerous faint, low-mass dE galaxies in the cluster core.

\section{Preliminary Aspects of our dE Spectroscopy}

The Keck I telescope and the Low-Resolution Imaging Spectrograph (LRIS)
were used in $\simeq 1.0$ arcsec seeing to obtain $2\times 1800$ sec
exposures of four fields, with $\simeq 22$ objects per field.  The
target objects were selected from the photometric sample described
above, based upon their position within the CMD, with the additional
constraint of slit placement.  Our wavelength coverage is $3800 \lae
\lambda \lae 6200$ \AA, with 8 \AA \ resolution and S/N $\simeq 50$
for the dEs.  We have preliminary results from an LRIS field located
$\simeq 3.8$ arcmin west of NGC 4874.  Of the 24 objects covered with
slits, 17 were ``targets'' (with $15.5 < R < 20.5$ mag and with $0.7 <
(B-R) < 1.9$ mag), the others chosen to fill remaining slitlets.
Based upon the number density of control-field objects in this region
of the CMD, we would expect $\simeq 9-10$ of these targets to be
genuine Coma members.  However, only four of these targets have Coma
velocities; a large fraction of the other targets are emission-line
galaxies at $z \simeq 0.2$.  If this unexpectedly-low fraction of Coma
members is not a result of small number statistics, and if it
continues to the other three LRIS fields, it will constraint
measurements of the Coma luminosity function which rely on correction
using associated control fields (Bernstein et al. 1995; Secker \&
Harris 1996).  Refer to Adami et al. (1997) for preliminary results of
velocities for objects in the Bernstein et al. (1995) field.  Using
the dE spectra, our primary goals are to estimate age and abundances
by measuring Lick spectral line indices (Worthey 1994), and to address
the following questions:

(a) What is the mean metallicity-luminosity relation determined from
metallicity-sensitive indices (e.g., Fe4668, Fe5709, and Fe5782); are
the high metallicities inferred by integrated colors confirmed by
spectroscopy?

(b) Is there evidence from age-sensitive indices (e.g., G4300,
$H\beta$ and the higher-order Balmer lines) that the dEs are composed
of a homogeneous population of old stars as in globular clusters, or
is there evidence for young and/or intermediate age components (as was
found in some Fornax dEs; Held \& Mould 1994)?

\section*{Acknowledgements} 

The W.M. Keck Observatory is operated as a scientific partnership
between the California Institute of Technology, the University of
California, and NASA.  It was made possible by the generous financial
support of the W.M. Keck Foundation.

\end{document}